\documentstyle[12pt,epsfig]{article}
\begin{document}
\noindent
\begin{center}
{\Large {\bf Cosmological Exact Solutions\\
in Some Modified Gravitational Theories}}\\ \vspace{2cm}
 ${\bf Yousef~Bisabr}$\footnote{e-mail:~y-bisabr@srttu.edu.}\\
\vspace{.5cm} {\small{Department of Physics, Shahid Rajaee Teacher
Training University,
Lavizan, Tehran 16788, Iran}}\\
\end{center}
\vspace{1cm}
\begin{abstract}
In a homogenous and isotropic cosmology, we introduce general exact solutions for some modified gravity models.  In particular, we introduce exact solutions for power-law $f(R)$ gravity and
Brans-Dicke theory in Einstein and Jordan conformal frames.   In the Brans-Dicke case, the solutions are presented for both single and double exponential potentials in Einstein
frame which correspond to power-law potentials in Jordan frame.  Our analysis for extracting general exact solutions can also be generalized to those scalar-tensor
theories in which the scalar field has an exponential coupling to Ricci scalar.

\end{abstract}

\section{Introduction}
General Relativity is a powerful tool to explain theoretically many observational facts about the Universe.  Despite all the successes, there are also some unresolved problems such as inflation, the cosmological constant problem and the problems associated with the dark sector, i.e., dark matter and dark energy.  These problems have motivated people to seek for some modifications of the theory.  Among many possibilities, much attention has been paid in recent years to two classes of modified gravity theories; scalar-tensor theories \cite{sc}
in which gravity is described not only by a metric tensor but also by a scalar field and $f(R)$ gravity theories \cite{I1} in which $R$ in the gravitational action is replaced by the function $f(R)$.\\
Comparing with General Relativity, these theories consider new degrees of freedom which open new possibilities for addressing the aforementioned problems. In scalar-tensor gravity, one introduces a new dynamical scalar field and in $f(R)$ gravity one considers forth order
derivatives of the metric in the field equations.  These modifications
make the theories admit a larger variety of solutions than Einstein equations
in General Relativity at the cost of
increased complexity.   This increased complexity together with nonlinearity make finding exact solutions for the field equations be much more difficult.\\
Despite the complexities, some cosmological exact solutions have been found in both classes.  In scalar-tensor gravity, exact solutions are usually given when
the scalar field has no potential function \cite{sc}.  There are also exact solutions for some functional forms of the potential such as power-law potential \cite{qu} and exponential potential when coupling of the scalar field is minimal \cite{ex}.  In power-law $f(R)$ gravity, exact solutions have been reported
in homogenous and isotropic cosmology in \cite{clif1} \cite{clif2}. In the present work we are looking for exact solutions in the two modified gravity models with a different approach.
We start with Einstein frame representation of a general scalar-tensor gravity in which there is an interaction between the scalar field and matter systems.  This representation is the same for all $f(R)$ gravities and all parameterizations of scalar-tensor theories.  Different gravitational theories are characterized by their potential, their coupling functions and the coupling
strengths ($A(\varphi)$ and $\beta(\varphi)$ in the following).  We introduce general exact solutions of the
 gravitational models for which the coupling strength takes constant values.\\
The plan of the paper is the following : In section 2, we first consider the action of a general scalar-tensor theory in Einstein frame in which the matter system is taken to be a
perfect fluid with a barotropic equation of state.  We then write the corresponding field equations in a homogenous and isotropic cosmology describing by the
Friedman-Robertson-walker (FRW) metric.  We assume spacetimes with zero spatial geometries.  The matter system and the scalar field are not separately conserved due to the interaction
and there is an energy flow between the two components.  Direction of the energy flow depends on the coupling strength and the equation of state of the fluid.  For an equation of state corresponding to radiation there is no interaction and hence no energy flow.  In section 3, we present the solutions of the field equations by direct integration.
We first solve the continuity equation for the matter system.  The solution indicates that evolution of the matter energy density is modified due to the interaction.  We can consider
this modification as a modification of the exponent of the scale factor by an arbitrary function $\epsilon$.  Here $\epsilon$ measures
the energy exchange between the matter and the scalar field.  This leads to a relation between the scalar field and the scale factor.  A relevant role is played by this relation
to reduce the second order differential equation of the scalar field to a first order one for obtaining the potential function.  We will show that exact solutions can be obtained when  $\epsilon$ and the coupling strength take constant values and for exponential forms of the potential.  We argue that this procedure can be applied to power-law $f(R)$ gravity
and some scalar-tensor gravities such as BD theory.  In section 4, we draw our conclusions.
~~~~~~~~~~~~~~~~~~~~~~~~~~~~~~~~~~~~~~~~~~~~~~~~~~~~~~~~~~~~~~~~~~~~~~~~~~~~~~~~~~~~~~~~~~\\\\

\section{Field Equations}
 We consider the action functional
 \begin{equation}
S_{EF}=\frac{1}{2} \int d^{4}x \sqrt{-g}~\{M_p^2
R-g^{\mu\nu} \nabla_{\mu} \varphi~ \nabla_{\nu} \varphi
-2V(\varphi)\}+ S_{m}(A^2(\varphi)g_{\mu\nu}
, \psi) \label{b4}\end{equation}
where $M_p^{-2}\equiv 8\pi G$, $G$ is
the gravitational constant and $S_{m}$ is the action of some matter
field $\psi$. The function $A(\varphi)$ is a coupling function that
characterizes coupling of the scalar field $\varphi$ with the matter
sector.  The action (\ref{b4}) may be taken as Einstein frame representation of a generalized scalar-tensor gravity
or an $f(R)$ gravity model .\\
Variation of this action with respect to the metric tensor $g^{\mu\nu}$,
leads to
\begin{equation}
G_{\mu\nu}=M_p^{-2}(T^{\varphi}_{\mu\nu}+
T^{m}_{\mu\nu}) \label{b6}
\label{b7}\end{equation} where
\begin{equation}
T^{\varphi}_{\mu\nu}=\nabla_{\mu}\varphi
\nabla_{\nu}\varphi-\frac{1}{2}g_{\mu\nu}\nabla^{\gamma}\varphi
\nabla_{\gamma}\varphi-V(\varphi)g_{\mu\nu}
\label{b8}\end{equation}
\begin{equation}
T^m_{\mu\nu}=\frac{-2}{\sqrt{-g}}\frac{\delta S_{m}(g_{\mu\nu}, \psi)}{\delta g^{\mu\nu}} \label{b9}\end{equation} are
stress-tensors of the scalar field and the matter system.
 Variation with respect to the scalar field $\varphi$, gives
\begin{equation}
\Box\varphi-\frac{dV(\varphi)}{d\varphi}=-\frac{\beta(\varphi)}{M_p} T^{m}
\label{b11}\end{equation}
where
\begin{equation}
\beta(\varphi)=M_p \frac{d\ln A(\varphi)}{d\varphi}
\label{b8-1}\end{equation}
and $T^m\equiv g^{\mu\nu}T^m_{\mu\nu}$.  The two
stress-tensors $T^m_{\mu\nu}$ and $T^{\varphi}_{\mu\nu}$
are not separately conserved.
 Instead they satisfy the following equations
\begin{equation}
\nabla^{\mu}T^{m}_{\mu\nu}=-\nabla^{\mu}T^{\varphi}_{\mu\nu}= \frac{\beta(\varphi)}{M_p}\nabla_{\nu}\varphi~T^{m}\label{b13}\end{equation}
We apply the field equations in a
spatially flat homogeneous and isotropic cosmology described by
FRW spacetime
\begin{equation}
ds^2=-dt^2+a^2(t)(dx^2+dy^2+dz^2)
\end{equation}
where $a(t)$ is the scale factor. To do this, we take
$T^m_{\mu\nu}$ as the stress-tensor of a perfect fluid with energy density
$\rho_{m}$ and pressure
$p_m$. In this
case, the gravitational equations (\ref{b7}) takes the form
\begin{equation}
3\frac{\dot{a}^2}{a^2}=M_p^{-2}(\rho_{m}+\rho_{\varphi})
\label{b14}\end{equation}
\begin{equation}
2\frac{\ddot{a}}{a}+\frac{\dot{a}^2}{a^2}=-M_p^{-2}(p_m+p_{\varphi})
\label{b14-1}\end{equation}
where
$\rho_{\varphi}=\frac{1}{2}\dot{\varphi}^2+V(\varphi)$ , $p_{\varphi}=\frac{1}{2}\dot{\varphi}^2-V(\varphi)$ and
overdot indicates differentiation with respect
to the cosmic time $t$.  Combining these equations, gives
\begin{equation}
\frac{\ddot{a}}{a}+2\frac{\dot{a}^2}{a^2}=M_p^{-2}[\frac{1}{2}(\rho_m-p_m)+V(\varphi)]
\label{b14-2}\end{equation}
From (\ref{b11}) and (\ref{b13}), we obtain
\begin{equation}
\ddot{\phi}+3\frac{\dot{a}}{a}\dot{\varphi}+\frac{dV(\varphi)}{d\varphi}=-\frac{\beta(\varphi)}{M_p} (\rho_{m}-3p_m)
\label{b15}\end{equation}
\begin{equation}
\dot{\rho}_{m}+3\frac{\dot{a}}{a}(\omega_m+1)\rho_{m}=Q \label{b17}\end{equation}
\begin{equation}
\dot{\rho}_{\phi}+3\frac{\dot{a}}{a}(\omega_{\varphi}+1)\rho_{\varphi}=-Q
\label{b18}\end{equation} where
\begin{equation}
Q=\frac{\beta(\varphi)}{M_p} \dot{\varphi}(\rho_{m}-3p_m)
\label{b-18}\end{equation}
is the interaction term, $\omega_m\equiv p_m/\rho_m$ and $\omega_{\varphi}\equiv p_{\varphi}/\rho_{\varphi}$.  The direction of energy
transfer depends on the sign of $Q$.  For
$Q>0$, the energy transfer is from the scalar field (dark energy\footnote{The action (\ref{b4}) and the subsequent field equations are similar to those of some
interacting models in which there is an interaction between dark energy and (dark) matter \cite{c2}.  In those models, the coupling function $A(\varphi)$ is
usually taken as a pre-assumed function that is specified, for instance, by phenomenological arguments.}) to the
matter system and for $Q<0$ the reverse is true.
~~~~~~~~~~~~~~~~~~~~~~~~~~~~~~~~~~~~~~~~~~~~~~~~~~~~~~~~~~~~~~~~~~~~~~~~~~~~~~~~
\section{Solutions}
There are three independent field equations among (\ref{b14})-(\ref{b18}) for finding $a(t)$, $\phi(t)$ and $\rho_m(t)$\footnote{Here we assume
that the perfect fluid that describes the matter system has a constant equation of state parameter $\omega_m$.}. The equation (\ref{b17}) can be easily solved
\begin{equation}
\rho_m=\rho_{0m}~a^{-3(\omega_m+1)}~e^{\frac{(1-3\omega_m)}{M_p}\int \beta d\varphi}
\label{c1}\end{equation}
in which $\rho_{0m}$ is an integration constant.  This solution indicates that the evolution of energy density is modified due to
interaction of $\varphi$ with matter.  There will be no loss of generality if we write (\ref{c1}) as
\begin{equation}
\rho_m=\rho_{0m}~a^{-3(\omega_m+1)+\epsilon}
\label{c3}\end{equation}
with $\epsilon$ being defined by
\begin{equation}
\epsilon=\frac{(1-3\omega_m)\int\beta d\varphi}{M_p\ln a}
\label{c2-1}\end{equation}
Even though $\epsilon $ and the coupling strength $\beta$ in (\ref{c2-1}) are generally evolving functions, we will restrict ourselves to the
case that they can
be regarded as constant parameters.  In this case, (\ref{c2-1}) reduces to
\begin{equation}
\varphi=\sigma M_p\ln a
\label{c2}\end{equation}
with $\sigma$ being a constant defined by the relation $\epsilon=\beta \sigma(1-3\omega_m)$.
The expression (\ref{c3}) is similar to the rule presented by some authors for
characterizing decaying law of vacuum energy into dark matter \cite{al}.  It states that when $\epsilon>0$, matter is
created and energy is constantly injecting into the matter so
that the latter will dilute more slowly compared to its
standard evolution $\rho_m\propto a^{-3(\omega_m+1)}$. Similarly, when $\epsilon<0$
the reverse is true, namely that matter is annihilated and
the direction of energy transfer is outside of the matter system so that the rate of dilution is faster than
the standard one.\\
We will show that (\ref{c2}) actually satisfies (\ref{b15}) for some potentials.  It is possible to find those potentials by solving a
first order differential
equation.  There are two important cases that the function $\beta(\varphi)$ takes a constant configuration; $f(R)$ gravity and
Brans-Dicke theory.  We consider the two cases separately in the following :
\subsection{$f(R)$ Gravity}
The action for an $f(R)$
gravity theory in Jordan frame is given by
\begin{equation}
S_{JF}= \frac{1}{2}M_p^2\int d^{4}x \sqrt{-\bar{g}} f(\bar{R}) +S_{m}(\bar{g}_{\mu\nu}, \psi)\label{b1}\end{equation} where
$\bar{g}_{\mu\nu}$ is the metric in Jordan frame.  We consider a conformal transformation
\begin{equation}
g_{\mu\nu} =A^{-2}(\varphi)~ \bar{g}_{\mu\nu} \label{b2}\end{equation}
with
$A^{-2}(\varphi)\equiv\frac{df}{dR}=f^{'}(R)$.  This together with
\begin{equation} \varphi = \frac{M_p}{\beta } \ln A(\varphi)
\label{b3}\end{equation}
and
$\beta=-\sqrt{\frac{1}{6}}$, transforms (\ref{b1}) into the action (\ref{b4}) with a potential \cite{soko} \cite{w}
\begin{equation}
V(\varphi(R))=\frac{M_p^2}{2}(\frac{R}{f'(R)}-\frac{f(R)}{f'^2(R)})
\label{b5}\end{equation}
Instead of solving the field equations for a given $f(R)$ function (or, equivalently, a given potential function), we are looking for those models that accept
the solution (\ref{c2}).  In this way, one can reduce the second order differential equation of $\varphi$ to a first order one for finding the functional form of
 $V(\varphi)$.  To do this, we first put (\ref{c2}) into (\ref{b15}) which leads to
\begin{equation}
\frac{\ddot{a}}{a}+2\frac{\dot{a}^2}{a^2}=-\frac{\beta}{\sigma M_p^2}(\rho_m-3p_m)-\frac{1}{\sigma M_p}\frac{dV(\varphi)}{d\varphi}
\label{c4}\end{equation}
Comparing the latter with (\ref{b14-2}) gives then a consistency relation
\begin{equation}
\frac{dV(\varphi)}{d\varphi}+\frac{\sigma}{M_p}V(\varphi)=\alpha\rho_{m}
\label{c5}\end{equation}
where $\alpha=\frac{\sigma}{2M_p}(\omega_m-1)+\frac{\beta}{M_p}(3\omega_m-1)$.  This together with (\ref{c3}) and (\ref{c2}) gives
\begin{equation}
\frac{dV(\varphi)}{d\varphi}+\frac{\sigma}{M_p}V(\varphi)=\alpha\rho_{0m}e^{[-3(\omega_m+1)+\epsilon]\varphi/\sigma M_p}
\label{c5-1}\end{equation}
This is a first order differential equation which leads to the following solution
\begin{equation}
V(\varphi)=C_1e^{-\sigma\varphi/M_p}+\delta \rho_{0m}e^{[-3(\omega_m+1)+\epsilon]\varphi/\sigma M_p}
\label{c6}\end{equation}
where
\begin{equation}
\delta=\frac{\frac{\sigma}{2}(\omega_m-1)+\beta(3\omega_m-1)}{[-3(\omega_m+1)+\beta\sigma(1-3\omega_m)]\frac{1}{\sigma}+\sigma}
\label{c7}\end{equation}
and $C_1$ is an integration constant.
Thus the expression (\ref{c2}) is an exact solution of the field equations for the
potential (\ref{c6}). For $C_1=0$, (\ref{c6}) is reduced to a single exponential
potential\footnote{The case $C_1\neq 0$ corresponds to a double exponential potential which will be considered
 later.}
\begin{equation}
V(\varphi)=\delta \rho_{0m}e^{[-3(\omega_m+1)+\epsilon]\varphi/\sigma M_p}
\label{c6-1}\end{equation}
Putting this potential into the Friedman equation (\ref{b14}), gives
\begin{equation}
\frac{\dot{a}^2}{a^2}=u~a^{-3(\omega_m+1)+\epsilon}
\label{c8}\end{equation}
where
\begin{equation}
u=\frac{\rho_{0m}}{M^2_p}\frac{\delta+1}{3-\sigma^2/2}
\label{c8-1}\end{equation}
For $\sigma^2\neq 6$, this integrates to
\begin{equation}
a(t)=\{\gamma (\sqrt{u}~t+C_2)\}^{1/\gamma}
\label{c9}\end{equation}
with $\gamma=\frac{1}{2}[3(\omega_m+1)-\epsilon]$.  For $\epsilon<0$, $\gamma$ remains positive for a perfect fluid which satisfies weak energy condition $\omega_m+1\geq 0$.  However when $\epsilon>0$, there is a chance for $\gamma<0$ and a contracting
scale factor.  This case happens when the energy flow from $\varphi$ to matter is so strong that make the Universe collapse.\\
To answer the question of how this solution is attributed to $f(R)$ function, we should use a back mapping of our
results to the Jordan frame.  One may use (\ref{c2}), (\ref{b2}) and (\ref{b3}) to write
\begin{equation}
\bar{a}(t)=A(\varphi)a(t)=e^{\frac{\beta\varphi}{M_p}} a(t)=a^{\beta\sigma+1}(t)=\{\gamma (\sqrt{u}~t+C_2)\}^{(\beta\sigma+1)/\gamma}
\label{c9-1}\end{equation}
$$
d\bar{t}=A(\varphi)~dt=e^{\frac{\beta\varphi}{M_p}}~ dt=a^{\beta\sigma}(t)~dt=\{\gamma
(\sqrt{u}~t+C_2)\}^{\beta\sigma/\gamma}~dt
$$
\begin{equation}
\Rightarrow\bar{t}=\{\gamma
(\sqrt{u}~t+C_2)\}^{\frac{\beta\sigma}\gamma+1}/\sqrt{u}(\beta\sigma+\gamma)~~~~~~~~~~~~~~~~~~~~~~~~~~~~~~~
\label{c9-2}\end{equation}
Combining the latter two relations gives $\bar{a}(\bar{t})$, scale
factor in the Jordan frame. For $C_2=0$, scale factor has a power-law form both in Einstein and Jordan frames :
$$
if ~~~\omega_m=1/3~~~~\Rightarrow~~~~~~~~~a(t)\propto
t^{\frac{1}{2}}~~~~~~~~~~,~~~~~~~~~~~~~~~~~~\bar{a}(\bar{t})\propto
\bar{t}^{\frac{\beta\sigma+1}{\beta\sigma+2}}
$$
\begin{equation}
if ~~~~\omega_m=0~~~~~\Rightarrow~~~~~~~~~~~a(t)\propto
t^{\frac{2}{3-\beta\sigma}}~~~~~~~~~,~~~~~~~~~~~~\bar{a}(\bar{t})\propto
\bar{t}^{\frac{2(\beta\sigma+1)}{(\beta\sigma+3)}}
\label{c9-7}\end{equation}
These solutions indicate that even hough passing from one conformal frame to the other does not change the power-law behavior, it changes the exponents.  However, in the case that
$\beta\sigma<<1$, the exponents of the scale factor follow the standard evolution in the radiation and the dust phases in both conformal frames.\\  The function $f(R)$ depends crucially on the functional form of the
potential $V(\varphi)$. For a single exponential potential of the form
(\ref{c6-1}), the expression (\ref{b5}) gives a differential equation
\begin{equation}
V(R)=\delta \rho_{0m}~f'^{\sqrt{\frac{3}{2}}[-3(\omega_m+1)+\epsilon]/\sigma}=\frac{M_p^2}{2}(\frac{R}{f'}-\frac{f}{f'^2})
\label{c9-3}\end{equation}
This has a simple power-law solution $f(R)=f_0~R^n$ with $n$ being
\begin{equation}
n=\frac{(\omega_m+1)(\beta\sigma+1)}{(\omega_m+1)+\beta\sigma(\omega_m+1/3)}
\label{c9-4}\end{equation}
and $f_0$ is given by
\begin{equation}
2\delta\frac{\rho_{0m}}{M_p^2}=(\frac{1}{nf_0})^{\frac{1}{n-1}}-f_0(\frac{1}{nf_0})^{\frac{n}{n-1}}
\label{c9-5}\end{equation}
Thus (\ref{c9})-(\ref{c9-7}) are exact solutions for power-law $f(R)$ gravity model.  It should also be noted that $n$ is given in terms of two free parameters
$\omega_m$ and $\sigma$ (or $\epsilon$) so that our solutions consider a much wider variety of solutions comparing with those obtained in \cite{clif2}.

\subsection{Scalar-Tensor Gravity}
The general action of a scalar-tensor gravity is given by \cite{pol}
\begin{equation}
S_{JF}=\frac{1}{16\pi G} \int d^{4}x \sqrt{-\bar{g}}~\{F(\phi)\bar{R}-Z(\phi)\bar{g}^{\mu\nu} \bar{\nabla}_{\mu} \phi~ \bar{\nabla}_{\nu} \phi
-2U(\phi)\}+ S_{m}( \bar{g}_{\mu\nu}, \psi_m) \label{d1}\end{equation}
where $F(\phi)$, $Z(\phi)$ and $U(\phi)$ are some functions\footnote{ One can always redefine the scalar field
to reduce $F(\phi)$ and $Z(\phi)$ to only one unknown function.}.  This action is reduced to the action (\ref{b4}) by
the conformal transformation (\ref{b2}) with $A(\varphi)=F^{-1/2}(\phi)$ and
\begin{equation}
(\frac{d\varphi}{d\phi})^2=2M_p^2[\frac{3}{4}(\frac{d\ln F(\phi)}{d\phi})^2+\frac{Z(\phi)}{2F(\phi)}] \label{d2}\end{equation}
\begin{equation}
V(\varphi)=M_p^2 U(\phi)F^{-2}(\phi)\label{d3}\end{equation}
The coupling function $A(\varphi)$ depends on the functions $F(\phi)$, $Z(\phi)$ and $U(\phi)$ through the relation (\ref{d2}).  For some
particular choices
of these functions, $\beta$ takes a constant configuration and then, as a result of (\ref{b8-1}), $A(\varphi)$ takes an exponential form.  These cases define a class of scalar-tensor
theories for which
the solution (\ref{c2}) and all the subsequent results obtained in the subsection $3.1$ are valid.  In the following, we will restrict ourselves to this class
of scalar-tensor theories.   One important theory in this class is given by the BD parameterization in which $F(\phi)=16\pi G\phi$, $Z(\phi)=16\pi G\omega_{BD}/\phi$
and $U(\phi)=8\pi G W(\phi)$, and then
\begin{equation}
S_{JF}= \int d^{4}x \sqrt{-\bar{g}} (\phi \bar{R} -\frac{\omega_{BD}}{\phi}\bar{g}^{\mu\nu}\bar{\nabla}_{\mu}\phi \bar{\nabla}_{\nu}\phi-W(\phi))+S_{m}(\bar{g}_{\mu\nu}, \psi)\label{d5}\end{equation}
with $\omega_{BD}$ and $W(\phi)$ being BD parameter and the potential in Jordan frame, respectively.  This action is reduced to (\ref{b4}) by \cite{sc} \cite{bis}
\begin{equation}
A(\varphi)=e^{\beta_{BD}\varphi/M_p}\label{d6}\end{equation}
\begin{equation}
\varphi(\phi)/M_p=\sqrt{\omega_{BD}+3/2}\ln (\frac{\phi}{\phi_0})
\label{a3}\end{equation}
\begin{equation}
V(\varphi)=W(\phi(\varphi))~e^{8\beta_{BD}\varphi/M_p}
\label{d6-1}\end{equation}
where $\phi_0\sim G^{-1}$ and $\beta_{BD}=-1/2\sqrt{\omega_{BD}+3/2}$.  The exact solutions of this BD model with potential (\ref{c6-1})
are given by (\ref{c9}) just by replacing $\beta$ with $\beta_{BD}$.  Note that when $\omega_{BD}\rightarrow 0$,
then $\beta_{BD}\rightarrow \frac{-1}{\sqrt{6}}$ and Einstein frame representations of BD model and $f(R)$ gravity are the same.  In Jordan frame, on the other hand,
the solutions are (\ref{c9-1})-(\ref{c9-7}) with $\beta$ being replaced by $\beta_{BD}$ and the corresponding potential is of power-law form
\begin{equation}
W(\phi)=\delta \rho_{0m}(\frac{\phi}{\phi_0})^{(4+\frac{\gamma}{\sigma\beta_{BD}})}
\label{d6-2}\end{equation}
The solutions are plotted in fig.1 for both $f(R)$ and
BD gravity models.  The figure indicates that the distinction between the two theories becomes more apparent when time grows.\\There is a large lower bound imposed on $\omega_{BD}$ by solar system
\begin{figure}[ht]
\begin{center}
\includegraphics[width=0.49\linewidth]{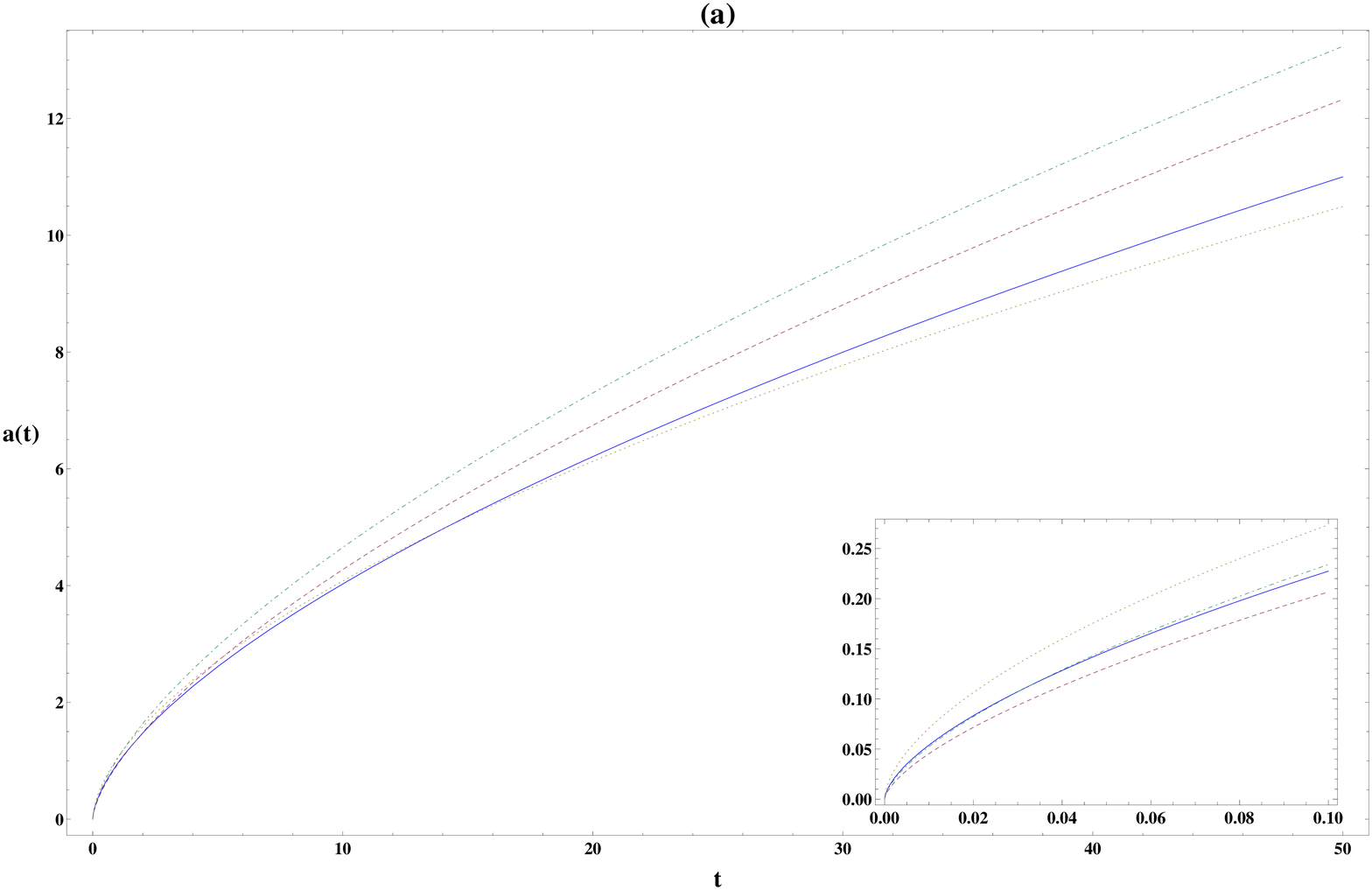}
\includegraphics[width=0.49\linewidth]{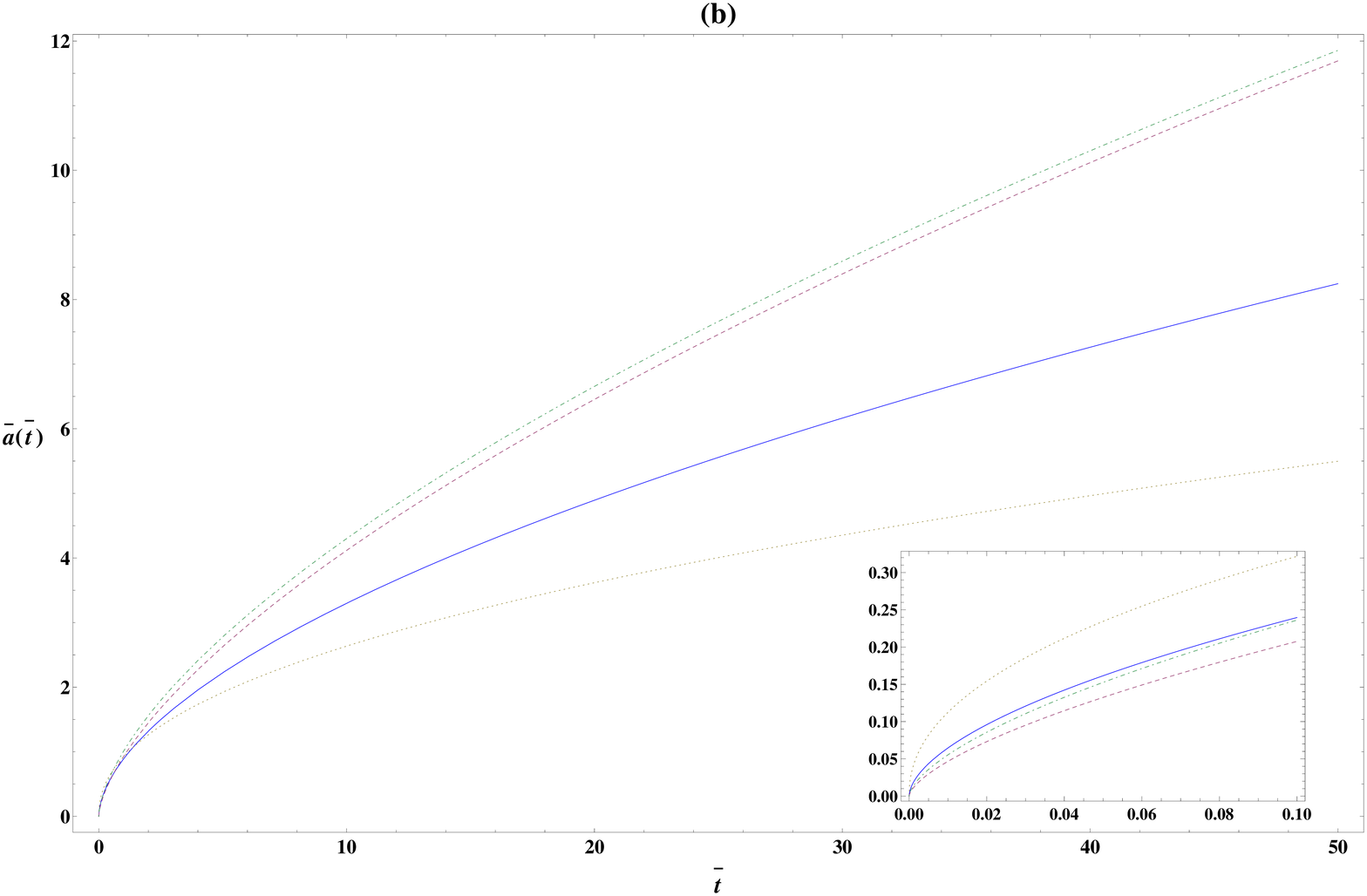}
\caption{The plot of scale factor for dust in (a) Einstein and (b) Jordan frames.  The solid and dashed lines correspond to $\sigma=0.5$ while
dotted and dashed-dotted lines correspond to $\sigma=1$ in $f(R)$ and BD gravity, respectively. }
\end{center}
\end{figure}
experiments \cite{omeg} which implies that $\beta_{BD}<<1$.   If $\sigma$ remains of order of unity, then $\beta_{BD}\sigma<<1$ and the exponents of the scale factor
in the power-law solutions (\ref{c9-7}) follow the standard evolution in radiation and matter phases both in Einstein and Jordan conformal frames.  \\
The condition for existing a late-time accelerating phase is given by $q=-\ddot{a}a/\dot{a}^2<0$ with $q$ being the deceleration parameter.  For the solution (\ref{c9}) in the dust phase, this condition gives $\sigma>1/\beta$ (in the BD case $\sigma>1/\beta_{BD}$).  Since $\beta_{BD}<\beta$ then accelerating solutions exist for larger values of the parameter $\sigma$ in the BD theory with respect to $f(R)$ gravity.\\
In potential (\ref{c6}), one can take $C_1\neq 0$ which leads to a double exponential potential.   The integration constant $C_1$
can be determined by noting the fact that when $\varphi$ takes a constant configuration in the action (\ref{b4}), then $V(\varphi)$ acts as a constant cosmological term. One may consider the condition
\begin{equation}
V(\varphi=0)=C_1+\delta \rho_{0m}=M_p^2\Lambda \equiv \rho_{\Lambda}
\label{c-66}\end{equation}
which results in $C_1=\rho_{\Lambda}-\delta\rho_{0m}$.  The relation (\ref{c6}) takes then the form
\begin{equation}
V(\varphi)/\rho_{0m}=(\rho_{\Lambda}/\rho_{0m}-\delta)e^{-\sigma\varphi/M_p}+\delta e^{[-3(\omega_m+1)+\epsilon]\varphi/\sigma M_p}
\label{c8}\end{equation}
For this potential function, the Friedmann equation (\ref{b14}) becomes
\begin{equation}
\frac{\dot{a}^2}{a^2}=u~ a^{-3(\omega_m+1)+\epsilon}+v~ a^{-\sigma^2}
\label{c9-9}\end{equation}
where
\begin{equation}
v=\frac{\rho_{m0}}{M_p^2}\frac{\rho_{\Lambda}/\rho_{m0}-\delta}{3-\sigma^2/2}
\label{c11}\end{equation}
For $\sigma^2\neq 6$, exact solutions can be found by a direct integration for different values of $\omega_m$.  We first consider some popular cases.\\
 $1)~ \omega_m=-1$ :\\
In this case, (\ref{c7}), (\ref{c8-1}) and (\ref{c11}) give  $\delta=-1$, $u=0$ and $v=\frac{\rho_{0m}}{M_p^2}\frac{10}{3(3-\sigma^2/2)}$ where we have set
$\rho_{\Lambda}/\rho_{0m}=7/3$ according to recent observations \cite{re}.  The solution of (\ref{c9-9}) is then
\begin{equation}
a(t)=[\frac{\sigma^2}{2}(\sqrt{v}~t+C_3)]^{2/\sigma^2}
\label{c12}\end{equation}
where $C_3$ is an integration constant.  The deceleration parameter is negative if $\sigma^2<2$ and the solution leads to a power-law inflation for a sufficiently small $\sigma$.\\
$2)~\omega_m=1/3$ :\\
In this case, $\delta=\frac{\sigma^2}{12-3\sigma^2}$, $u=\frac{\rho_{0m}}{M_p^2}\frac{4}{3(4-\sigma^2)}$ and $v=\frac{\rho_{0m}}{M_p^2}\frac{4(7-2\sigma^2)}{3(4-\sigma^2)(3-\sigma^2/2)}$.  We have simple solutions in some special cases
\begin{equation}
a(t)=[(t+C_4)^2-\frac{2}{3}]^{1/2}~~~~~~~~~~~~~~~~~~~~~~~~~~~~if~~~~~~~~~~~~~~~~~~~\sigma^2=2
\end{equation}
\begin{equation}
a(t)=[4\sqrt{2/3 }~t+2C_5]^{1/2}~~~~~~~~~~~~~~~~~~~~~~~~~~~if~~~~~~~~~~~~~~~~~~~~~\sigma^2=7/2
\end{equation}
The case $\sigma^2=7/2$ and $C_5=0$ corresponds to the standard Friedmann model $a(t)\propto t^{1/2}$ in the radiation phase.  The solution for any $\sigma$ is
$$
\frac{1}{\sigma^2(ua^{\sigma^2}+va^4)}\{2a^{3+\sigma^2}\sqrt{\frac{u}{a^2}+va^{2-\sigma^2}}\sqrt{1+\frac{ua^{\sigma^2-4}}{v}}2F1[\frac{1}{2},\frac{\sigma^2}{2(\sigma^2-4)},
1+\frac{\sigma^2}{2(\sigma^2-4)},-\frac{ua^{\sigma^2-4}}{v}]\}
$$
\begin{equation}
=t+C_6~~~~~~~~~~~~~~~~~~~~~~~~~~~~~~~~~~~~~~~~~~~~~~~~~~~~~~~~~~~~~~~~~~~~~~~~~~~~~~~~~~~~~~~~~~~~~~~~~~~~~~~~~~~~~~~~~~~~~~~~~~~~~~~~~~~~~~~~~~~~~~~~~~~
\label{c13}\end{equation}
where $2F1[a,b,c,x]$ is the hypergeometric function $_2F_1[a,b;c;x]$.  Note that this expression gives $t=t(a$).\\
$3)~\omega_m=0 :$\\
In this case, $\delta=\frac{-(\beta_{BD}+\sigma/2)}{(\beta_{BD}-3/\sigma)+\sigma}$ and $\epsilon_{BD}=\beta_{BD}\sigma$ and the solution of (\ref{c9-9}) is
$$
\frac{1}{\sigma^2(ua^{\sigma(\beta_{BD}+\sigma)}+va^3)}\{2a^{2+\sigma^2}\sqrt{ua^{\beta_{BD}\sigma-1}+va^{2-\sigma^2}}\sqrt{1+\frac{ua^{-3+\beta_{BD}\sigma+\sigma^2}}{v}}2F1[\frac{1}{2},
\frac{\sigma^2}{2(-3+\beta_{BD}\sigma+\sigma^2)},
$$
\begin{equation}
1+\frac{\sigma^2}{2(-3+\beta_{BD}\sigma+\sigma^2)},-\frac{ua^{-3+\beta_{BD}\sigma+\sigma^2}}{v}]\}
=t+C_7~~~~~~~~~~~~~~~~~~~~~~~~~~~~~~~~~~~~~~~~~~~~~~~~~~~~~~~~~~~~~~~~~~~~~~~~~~~~~~~~~~~~~~~~~~~~~~~~~~~~~~~~~~~~~~~~
\label{c14}\end{equation}
This solution is plotted in fig.2 for some values of the parameter $\sigma$ and $C_7=0$.
\begin{figure}[ht]
\begin{center}
\includegraphics[width=0.5\linewidth]{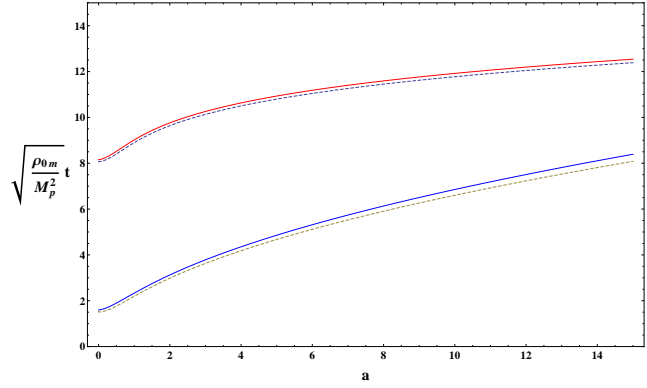}
\caption{The Plot of $t=t(a)$ in the dust region (the solution (\ref{c14}) for $C_7=0$).  The solid lines correspond to BD theory for $\beta_{BD}=-0.08$ (or $\omega_{BD}=4000$) and $\sigma=0.5, 1$ from top to bottom.  The dashed lines indicate $f(R)$ gravity ($\beta=-1/\sqrt{6}$) for the same values of the parameter $\sigma$.}
\end{center}
\end{figure}
The general solution of (\ref{c9-9}) for any $\sigma$ and $\omega_m$ is given by
$$
\frac{-1}{[\epsilon_{BD}-3(\omega_m+1)][ua^{\sigma^2+\epsilon_{BD}}+va^{3(\omega_{m}+1)}]}\{2a^{\sigma^2+(2+3\omega_m)}\sqrt{u a^{\epsilon_{BD}-(3\omega_{m}+1)}+v a^{-\sigma^2+2}}
$$
$$
\sqrt{1+\frac{v a^{-\sigma^2-\epsilon_{BD}+3(\omega_m+1)}}{u}}2F1[\frac{1}{2},\frac{\epsilon_{BD}-3(\omega_m+1)}{2[\sigma^2+\epsilon_{BD}-3(\omega_m+1)]},
1+\frac{\epsilon_{BD}-3(\omega_m+1)}{2[\sigma^2+\epsilon_{BD}-3(\omega_m+1)]},
$$
\begin{equation}
-\frac{v a^{-\sigma^2-\epsilon_{BD}+3(\omega_m+1)}}{u}]\}=t+C_8~~~~~~~~~~~~~~~~~~~~~~~~~~~~~~~~~~~~~~~~~~~~~~~~~~~~~~~~~~~~~~~~~~~~~~~
\label{c15}\end{equation}
where $\epsilon_{BD}=\beta_{BD} \sigma(1-3\omega_m)$.  It should be remarked that the solutions (\ref{c12})-(\ref{c15}) can also be regarded as the solutions of Einstein frame representation of an $f(R)$ gravity with potential function (\ref{c8}) just by replacing $\beta_{BD}$ and $\epsilon_{BD}$ with $\beta$ and $\epsilon$, respectively.  The $f(R)$ function which corresponds to this double exponential potential can be formally obtained by using (\ref{b5}) although it is not easy to solve the resulting differential equation for general $\sigma$ and $\omega_m$.  However, it is shown that \cite{kor} the Einstein frame representation of the models\footnote{These models have been investigated in \cite{ca} to explain the late-time acceleration of the  Universe.} $f(R)=R-\mu^{2(n+1)}/R^n$ has a potential function which takes a double exponential form when $n\rightarrow \infty$.  This potential corresponds to (\ref{c6}) in some regions of the parameters space.\\
The fig.2 plots the solution (\ref{c14}) for the two gravity models.  It indicates that the overall behavior of the solutions is the same for given values of the parameter $\sigma$.  It is important to note that there is no distinction between the two gravity models in the vacuum and the radiation phases since the solutions (\ref{c12})-(\ref{c13}) do not depend on the parameter $\beta$.\\
We may consider different scalar-tensor
gravity models for which (\ref{b8-1}) takes a constant value and then the coupling function $A(\varphi)$ becomes exponential.  Besides BD theory, in all these
cases (\ref{c2}) satisfies the field equations for some potential functions of the forms (\ref{c6-1}) and (\ref{c8}).  For instance, consider
$F(\phi)=F_0 e^{\eta\phi}$ and $Z(\phi)=Z_0 e^{\eta\phi}$ for which (\ref{d2}) gives $\varphi=\sqrt{2}\zeta M_p \phi$ with $\zeta=\sqrt{\frac{3}{4}\eta^2+\frac{Z_0}{2F_0}}$.  Then (\ref{b8-1}) results in
the coupling function $A(\varphi)=F_0^{-1/2} e^{\beta_c\varphi/M_p}$ in which $\beta_c\equiv-\frac{\eta}{2\sqrt{2}\zeta}$ is a constant.  This parameterization has been already studied
as extended quintessence model where the scalar
field playing the role of the dark energy is exponentially coupled to the Ricci scalar \cite{man}.  For the potentials (\ref{c6-1}) and (\ref{c8}), this scalar-tensor theory has general exact solutions of the forms (\ref{c9})-(\ref{c9-7}) and (\ref{c12})-(\ref{c15}) in which ($\beta$, $\epsilon$) and ($\beta_{BD}$, $\epsilon_{BD}$) should be replaced by ($\beta_c$, $\epsilon_c$).
\section{Conclusion}
We have introduced homogenous and isotropic exact cosmological solutions in some modified gravitational theories.  We assume that the Universe has a flat spatial geometry and is
filled with a perfect fluid with a barotropic equation of state.
In a general
scalar-tensor theory in Einstein frame, there is a coupling between the scalar field and the matter part through the coupling function $A(\varphi)$.  We have found general exact solutions for gravitational theories in which the coupling strength $\beta(\varphi)$ takes constant values. There are two groups of gravitational theories with a constant coupling strength, $f(R)$ gravity and a class of scalar-tensor gravity including BD theory. Exact solutions of these gravitational theories are found for single and double exponential potentials.   \\
For single exponential potentials,
the exact solutions are first presented in the Einstein frame for $f(R)$ gravity and BD theory and then transformed back the solutions to the Jordan frame.  It is shown that the solutions in Jordan frame belong to a power-law $f(R)$ gravity and BD theory with a power-law potential.  Analysis of the solutions indicates that there are power-law solutions for the scale factor  in Einstein and Jordan conformal frames for both gravitational theories.  When $\beta\sigma<<1$ ($\beta_{BD}\sigma<<1$) they reduce to the standard evolution in radiation and dust phases.
\\ For double exponential potentials, general exact solutions are also presented for both BD and $f(R)$ theories in Einstein frame.  It is interesting that these solutions do not distinguish between the two theories in vacuum and radiation
phases.  The analysis can be generalized to scalar-tensor theories in which the scalar field has an exponential coupling to Ricci scalar.  In this case, functional
form of the solutions are exactly the same as the solutions in the other two gravitational theories for single and double exponential potentials and can be obtained by usual replacement of the parameters with ($\beta_c$, $\epsilon_c$).\\\\\\

{\bf Acknowledgment}\\\\
I would like to thank Dr. R. Rashidi for useful discussions.

\end{document}